\documentclass[lettersize,journal]{IEEEtran}
\IEEEoverridecommandlockouts
\usepackage[table,dvipsnames]{xcolor}
\usepackage{array,amsthm}
\usepackage{subfigure,graphicx}
\usepackage{amsmath,amsfonts,amssymb,mathtools}
\usepackage{cite}
\usepackage{textcomp}
\usepackage{stfloats}
\usepackage{booktabs}
\usepackage{enumerate}
\usepackage{pgfplots}
\pgfplotsset{compat=1.18}
\usepackage{balance}
\usepackage[printonlyused,withpage]{acronym}
\usepackage[ruled,vlined,linesnumbered]{algorithm2e}
\SetKwInOut{Input}{Input}
\SetKwInOut{Output}{Output}
\usepackage[
colorlinks=true,
citecolor=blue,
linkcolor=blue,
urlcolor=black
]{hyperref}

\usepackage{tikz}
\usepackage{pifont}

\acrodef{KPI}[KPI]{key performance indicator}
\acrodef{WRT}[w.r.t.]{with respect to}
\acrodef{WP}[w.p.]{with probability}
\acrodef{LHS}[l.h.s.]{left hand side}
\acrodef{RHS}[r.h.s.]{right hand side}

\acrodef{2D}[2D]{two-dimensional}
\acrodef{3D}[3D]{three-dimensional}
\acrodef{3GPP}[3GPP]{3rd Generation Partnership Project}
\acrodef{AOA}[AOA]{angle-of-arrival}
\acrodef{AWGN}[AWGN]{additive white Gaussian noise}
\acrodef{BS}[BS]{base station}
\acrodef{CFR}[CFR]{channel frequency response}
\acrodef{CIR}[CIR]{channel impulse response}
\acrodef{CTRV}[CTRV]{\textit{Constant Turn Rate and Velocity}}
\acrodef{ELAA}[ELAA]{extremely large aperture array}
\acrodef{FF}[FF]{far-field}
\acrodef{ISAC}[ISAC]{integrated sensing and communication}
\acrodef{KF}[KF]{Kalman filter}
\acrodef{LMS}[LMS]{least means square}

\acrodef{LOS}[LOS]{line-of-sight}
\acrodef{LS}[LS]{least squares}
\acrodef{MAP}[MAP]{maximum a posteriori}
\acrodef{MC}[MC]{Monte Carlo}
\acrodef{MDP}[MDP]{Markov decision process}
\acrodef{MIMO}[MIMO]{multiple-input multiple-output}
\acrodef{ML}[ML]{maximum likelihood}
\acrodef{MLE}[MLE]{maximum likelihood estimation}
\acrodef{MMSE}[MMSE]{minimum mean square error}
\acrodef{MSE}[MSE]{mean square error}
\acrodef{MUI}[MUI]{multi-user interference}
\acrodef{MUSIC}[MUSIC]{multiple signal classification}
\acrodef{NF}[NF]{near-field}
\acrodef{NLOS}[NLOS]{non-line-of-sight}
\acrodef{NN}[NN]{neural network}
\acrodef{PDP}[PDP]{power delay profile}
\acrodef{RCS}[RCS]{radar cross section}
\acrodef{RF}[RF]{radio frequency}
\acrodef{RFID}[RFID]{radio frequency identification}
\acrodef{RFS}[RFS]{random finite set}
\acrodef{RI}[RI]{range information}
\acrodef{RL}[RL]{reinforcement learning}
\acrodef{RMSE}[RMSE]{root-mean-square error}
\acrodef{ReNN}[regression NN]{regression neural network}
\acrodef{ROI}[ROI]{region of interest}
\acrodef{RTT}[RTT]{round-trip time}
\acrodef{RV}[RV]{random variable}
\acrodef{SI}[SI]{soft information}
\acrodef{SIR}[SIR]{signal-to-interference ratio}
\acrodef{SLAM}[SLAM]{simultaneous localization and mapping}
\acrodef{SMC}[SMC]{sequential monte carlo}
\acrodef{SNR}[SNR]{signal-to-noise ratio}
\acrodef{SINR}[SINR]{signal-to-interference-plus-noise ratio}
\acrodef{TDOA}[TDOA]{time difference-of-arrival}
\acrodef{TOA}[TOA]{time-of-arrival}
\acrodef{TOF}[TOF]{time-of-flight}
\acrodef{UAV}[UAV]{unmanned aerial vehicles}
\acrodef{UE}[UE]{user equipment}
\acrodef{UKF}[UKF]{unscented Kalman filter}
\acrodef{UWB}[UWB]{ultra-wideband}
\acrodef{WLS}[WLS]{weighted least squares}
\acrodef{AI}[AI]{artificial intelligence}

\begin{document}

\title{Environment-aware Near-field UE Tracking under\\Partial Blockage and Reflection}

\author{
\vspace{0.5em}
Hyunwoo~Park,~\IEEEmembership{Graduate~Student~Member,~IEEE},
Hyeon Seok Rou,~\IEEEmembership{Member,~IEEE},\\
Giuseppe Thadeu Freitas de Abreu,~\IEEEmembership{Senior~Member,~IEEE}, and
Sunwoo~Kim,~\IEEEmembership{Senior~Member,~IEEE}\\
\vspace{-1.4em}
\thanks{Manuscript received month day, 2026; revised month day, year.
The fundamental research described in this paper was supported
by the National Research Foundation of Korea under Grant NRF2023R1A2C3002890.
\\\textit{(Corresponding author: Sunwoo Kim)}}%
\thanks{Hyunwoo Park and Sunwoo Kim are with the Department of Electronic Engineering, Hanyang University, Seoul 04763, South Korea \\(e-mail: stark95@hanyang.ac.kr; remero@hanyang.ac.kr).}
\thanks{Hyeon Seok Rou and Giuseppe Thadeu Freitas de Abreu are with the School of Computer Science and Engineering, Constructor University Bremen, 28759 Bremen, Germany (e-mail: hrou@constructor.university; gabreu@constructor.university).}
\vspace{-2ex}
}



\maketitle

\begin{abstract}
This paper proposes an environment-aware \ac{NF} \ac{UE} tracking method for extremely large aperture arrays. 
By integrating known surface geometries and tracking the \ac{LOS} and \ac{NLOS} indicators per antenna element, the method captures partial blockages and reflections specific to the \ac{NF} spherical-wavefront regime, which are unavailable under the conventional \ac{FF} assumption.
The \ac{UE} positions are tracked by maximizing the cosine similarity between the predicted and received channels, enabling tracking even under complete \ac{LOS} obstruction. 
Simulation results confirm that increasing environment-awareness improves accuracy, and that \ac{NF} consistently outperforms FF baselines, achieving a $0.22\,\mathrm{m}$ root-mean-square error with full environment-awareness.
\end{abstract}

\vspace{-1ex}
\begin{IEEEkeywords}
Environment-awareness, near-field, tracking, partial blockage, partial reflection.
\vspace{-1ex}
\end{IEEEkeywords}

\acresetall

\section{Introduction}
\vspace{-0.5ex}

\IEEEPARstart{N}{ext-generation} wireless networks are envisioned not merely as communication pipelines but as perceptive infrastructures that support physical \ac{AI} systems such as autonomous vehicles, drones, and robots~\cite{6Gvision, physical_AI}.
\Ac{ISAC} serves as a cornerstone of this vision, enabling the network to simultaneously transfer data and extract physical-world information from radio signals~\cite{ISAC_tutorial}.

Among diverse sensing functionalities, accurate \ac{UE} positioning stands out as a fundamental enabler of location-aware resource management, predictive beam alignment, and autonomous mobility support~\cite{9781656,loc_survey}.
In practice, however, reliable positioning is hindered by the complex propagation environment, where surrounding surfaces cause signal blockage and multipath reflections~\cite{10123922}.

In view of such challenging radio environments, the rapid advancement of digital twin technologies has opened a promising avenue by providing detailed environment geometry including the locations, orientations, and material properties of surrounding surfaces~\cite{9711524, DT_wireless, 8477101}.
Leveraging such information, several recent studies have proposed localization methods that predict and incorporate the multipath propagation via ray-tracing, voxelation, and radio fingerprinting, turning reflected paths into useful position-related observations~\cite{10918620, Rou_3dimaging_24, 10570572}.

Nevertheless, these approaches predominantly rely on \ac{FF} channel models and assume that intermediate parameters such as \ac{TOA} and \ac{AOA} have already been extracted prior to position estimation~\cite{8515231}, inherently limiting their applicability since the extraction step itself requires accurate path-level modeling that becomes increasingly difficult under complex propagation conditions.

Amidst these bottlenecks, a key trend emerged which fundamentally reshaped the propagation landscape -- the deployment of \acp{ELAA} at high-frequency bands, substantially extending the \ac{NF} region governed by the spherical wavefront model~\cite{NF_tutorial, 9733790, iimori2022grant}.
In the \ac{NF} regime, a single surface may block or reflect the signal for only a subset of antenna elements, giving rise to partial blockage and partial reflection unique to spherical-wavefront propagation~\cite{10817348, 10942642}.
These spatially-selective effects violate the all-or-nothing path model inherent to \ac{FF} channel parameter estimation, rendering existing environment-aware methods inadequate for \ac{NF} systems.

Recent \ac{NF} localization and tracking works have begun to account for such partial effects~\cite{9508850, 10706285,10920706}, however, they do not incorporate environment geometry and thus cannot predict reflection paths or model blockage from surfaces outside the direct \ac{LOS} path.
Conversely, existing environment-aware approaches~\cite{10918620, 10570572, 8515231} rely on \ac{FF} models that inherently cannot capture per-element partial effects.
A method that jointly exploits environment geometry and \ac{NF} propagation characteristics therefore remains unexplored.

\IEEEpubidadjcol

In light of the above, this paper addresses the aforementioned gap by proposing an environment-aware \ac{NF} \ac{UE} tracking method that directly estimates the \ac{UE} position from the received signal, bypassing conventional channel parameter extraction.
To the best of the authors' knowledge, this is the first work to jointly embed known surface geometries into a per-antenna-element \ac{NF} channel prediction framework for \ac{UE} tracking.
The main contributions are summarized as follows: \vspace{-0.25ex}

\begin{itemize}
\item A per-element channel prediction framework is developed which incorporates surface geometries to determine individual \ac{LOS} and \acs{NLOS} conditions. This captures partial blockage and reflection effects unique to the \ac{NF} spherical-wavefront regime, which are unmodeled under conventional \ac{FF} planar-wavefront assumptions.
\item A cosine-similarity-based estimator is formulated, which is invariant to unknown transmit power and pilot phase. By exploiting predicted reflections from known surfaces, the method enables continuous \ac{UE} tracking even under complete \ac{LOS} obstruction.
\end{itemize}

\textit{Notations:} Vectors and matrices are written in lower- and upper-
case bold letters respectively and sets by calligraphic font. For example, a vector, a matrix, and a set are denoted by $\mathbf{a}$, $\mathbf{A}$ and $\mathcal{A}$, respectively.
Superscripts $(\cdot)^\mathrm{H}$ denotes Hermitian transpose, $\imath=\sqrt{-1}$ is the imaginary unit, and $\lVert\cdot\rVert$ returns the Euclidean norm.

\section{UE Tracking System and Problem Formulation}\label{sys}

\subsection{UE Tracking System}\label{sys_1}

Consider a scenario in which a mobile single-antenna \ac{UE} communicates with an \ac{ELAA} \ac{BS} equipped with $N$ antennas.
A \ac{2D} environment is considered, containing $S$ surfaces that partially reflect or block signal paths -- each modeled as a line segment\footnote{The proposed per-element path evaluation extends directly to \ac{3D} environments by replacing line-segment intersection tests with ray-polygon tests at the same asymptotic complexity.}.
The \ac{BS} is assumed to have knowledge of a subset (up to a complete set) of these surfaces and tracks the \ac{UE}'s position by receiving its uplink signals.

Let $\mathcal{S}$ denote the set of all $S$ surfaces and $\mathcal{S}_{\mathrm{a}}\subseteq\mathcal{S}$ denote the subset of $S_{\mathrm{a}}$ surfaces known to the \ac{BS}.
The environment-awareness level is defined as the ratio $\eta = {S_{\mathrm{a}} / S} \in [0,1]$, representing the fraction of surfaces whose geometry is available at the \ac{BS}.

At time $k$, the uplink signal $\mathbf{z}_{k,m}\in\mathbb{C}^{N\times 1}$ received at the BS on the $m$-th subcarrier is expressed as
\begin{equation}
\mathbf{z}_{k,m} = \sqrt{\frac{P}{M}}\, \mathbf{h}_{m}(\mathbf{p}_k) \,x_{k,m} \exp(\imath \phi_{k,m}) + \mathbf{w}_{k,m}
\end{equation}
where $P$ is the transmit power of the UE, $M$ is the number of subcarriers, $\mathbf{h}_{m}(\mathbf{p}_k)\in\mathbb{C}^{N\times 1}$ denotes the channel vector from the \ac{UE} at position $\mathbf{p}_k$ to the \ac{BS}, $x_{k,m}$ is the unit-power transmitted pilot symbol, $\phi_{k,m}$ is the phase error caused by imperfect synchronization between the \ac{UE} and the \ac{BS}, and $\mathbf{w}_{k,m}\sim\mathcal{CN}\left(\mathbf{0}_{N},\sigma_{\mathrm{w}}^2\mathbf{I}_{N}\right)$ represents the \ac{AWGN}.
The noise variance per subcarrier is $\sigma_{\mathrm{w}}^2=k_\mathrm{B}T\Delta_{\mathrm{f}}F$, where $k_\mathrm{B}$ is the Boltzmann constant, $T$ the ambient temperature, $\Delta_{\mathrm{f}}$ the subcarrier spacing, and $F$ the noise figure.

The channel vector $\mathbf{h}_{m}(\mathbf{p}_k)$ will capture both \ac{LOS} and \ac{NLOS} propagation paths, including reflections from surrounding surfaces.
For tractability, this paper considers specular reflections from smooth surfaces, where the reflection coefficient follows the Fresnel equations~\cite{rappaport2010wireless}.
Considering single-bounce propagation, which is well justified since higher-order reflections are negligible relative to the noise floor at the considered frequency~\cite{38901}, the set of paths reaching the $n$-th antenna element is characterized using $\mathcal{S}$.

Next, define the \ac{LOS} path indicator for the $n$-th antenna element as
\begin{equation}
b_{n}^{\mathrm{LOS}}(\mathbf{p}\!:\!\mathcal{S}) = \begin{cases}
1 & \text{if } \overline{\mathbf{p}_{n}'\mathbf{p}}\cap s = \emptyset,\;\forall\, s\in\mathcal{S} \\
0 & \text{otherwise}
\end{cases}\label{blos}
\end{equation}
where $\overline{\mathbf{p}_{n}'\mathbf{p}}$ denotes the line segment connecting the $n$-th antenna position $\mathbf{p}_{n}'$ and the \ac{UE} position $\mathbf{p}$, in other words, the indicator $b_{n}^{\mathrm{LOS}}$ returns a value of $1$ only if the \ac{LOS} path between $\mathbf{p}_{n}'$ and $\mathbf{p}$ does not intersect with any surface in $\mathcal{S}$.

Similarly, each surface $s\in\mathcal{S}$ can potentially generate a reflection path.
The reflection point $\mathbf{r}_{n,s}(\mathbf{p})$ on surface $s$ for the $n$-th antenna is computed via the image method \cite{7152831}, and the resulting path is considered valid if \textit{(i)} the reflection point lies within the finite extent of surface $s$, and \textit{(ii)} neither the incident nor the reflected segment is blocked by any other surface in $\mathcal{S}$.
Then, \ac{NLOS} path indicator for surface $s$ is defined as
\begin{equation}
b_{n,s}^{\mathrm{NLOS}}(\mathbf{p}\!:\!\mathcal{S}) = \begin{cases}
1 & \text{if conditions \textit{(i)} and \textit{(ii)} hold,} \\
0 & \text{otherwise.}
\end{cases}\label{bnlos}
\end{equation}

In all, the set of propagation paths for the $n$-th antenna is given by
\begin{equation}
\mathcal{L}_{n}(\mathbf{p}\!:\!\mathcal{S}) = \bigl\{0 \,\text{ if }\, b_{n}^{\mathrm{LOS}}=1\bigr\}\cup\bigl\{s\in\mathcal{S} \,\text{ if }\, b_{n,s}^{\mathrm{NLOS}}=1\bigr\}\label{pathset}
\end{equation}
where the index $0$ denotes the \ac{LOS} path.
Note that both the number of propagation paths $|\mathcal{L}_n|$ and the positions of reflection points $\mathbf{r}_{n,s}(\mathbf{p})$ differ for each antenna element due to \ac{NF} effects, which induce partial blockage and partial reflection.
Furthermore, the $n$-th element of the channel vector is expressed as \vspace{-1ex}
\begin{equation}
h_{m,n}(\mathbf{p})=\!\!\!\!\!\sum_{l\in\mathcal{L}_{n}(\mathbf{p}:\mathcal{S})}\!\frac{\beta_{m,l}(\mathbf{p})}{\sqrt{\alpha_{m,n,l}(\mathbf{p})}}\exp\!{\Bigl(-\imath\frac{2\pi}{\lambda_m}d_{n,l}(\mathbf{p})\Bigr)}\label{hmn}
\end{equation}
where $\beta_{m,l}(\mathbf{p})\in[0,1]$ is the reflection coefficient with $\beta_{m,0}(\mathbf{p})=1$ for \ac{LOS}, and $\alpha_{m,n,l}(\mathbf{p})$ is the path loss.

The path loss follows~\cite{38901} \vspace{-1ex}
\begin{equation}
\alpha_{m,n,l}(\mathbf{p})=\Bigl(\frac{4\pi}{\lambda_m}d_{n,l}(\mathbf{p})\Bigr)^2\label{alpha}
\end{equation}
where $\lambda_m$ is the wavelength for the $m$-th subcarrier and $d_{n,l}(\mathbf{p})$ is the total propagation distance of the $l$-th path associated with the $n$-th antenna element and position $\mathbf{p}$.

The propagation distance $d_{n,l}(\mathbf{p})$ is expressed as \vspace{-0.5ex}
\begin{align}
&d_{n,l}(\mathbf{p})\label{dnl}\\
&=\begin{cases}
\lVert\mathbf{p}_{n}'-\mathbf{p}\rVert & \text{if } l=0\;\text{(\ac{LOS})} \\
\lVert\mathbf{p}_{n}'-\mathbf{r}_{n,l}(\mathbf{p})\rVert+\lVert\mathbf{r}_{n,l}(\mathbf{p})-\mathbf{p}\rVert & \text{if } l=s\;\text{(\ac{NLOS})}
\end{cases}. \nonumber \vspace{-1ex}
\end{align}

\vspace{-1ex}
\subsection{Problem Formulation}

The objective of the presented problem is to track the \ac{UE} position by exploiting environment geometry information alongside the received uplink signals.
Given knowledge of the surface subset $\mathcal{S}_{\mathrm{a}}$, the propagation paths can be predicted for any candidate \ac{UE} position $\mathbf{p}$ by evaluating~\eqref{blos}--\eqref{hmn} with $\mathcal{S}_{\mathrm{a}}$ in place of $\mathcal{S}$.
The predicted channel vector $\hat{\mathbf{h}}_{m}(\mathbf{p}\!:\!\mathcal{S}_{\mathrm{a}})\in\mathbb{C}^{N\times 1}$ whose $n$-th element is expressed as
\begin{align}
\hat{h}_{m,n}(\mathbf{p}\!:\!\mathcal{S}_{\mathrm{a}})=\!\!\sum_{l\in\mathcal{L}_{n}(\mathbf{p}:\mathcal{S}_{\mathrm{a}})}\frac{\beta_{m,l}(\mathbf{p})}{\sqrt{\alpha_{m,n,l}(\mathbf{p})}}\exp{\Bigl(-\imath\frac{2\pi}{\lambda_m}d_{n,l}(\mathbf{p})\Bigr)} \nonumber \\[-2.5ex] \label{hhat}
\end{align}
which shares the same functional form as the true channel in~\eqref{hmn} but is evaluated over the path set $\mathcal{L}_{n}(\mathbf{p}\!:\!\mathcal{S}_{\mathrm{a}})$ determined by the known surfaces only.
Trivially, when $\eta=1$ (i.e., $\mathcal{S}_{\mathrm{a}}=\mathcal{S}$), the predicted path set coincides with the true path set, such that $\hat{\mathbf{h}}_{m}(\mathbf{p}\!:\!\mathcal{S})=\mathbf{h}_{m}(\mathbf{p})$.
\newpage

A tracking scenario is considered in which the initial \ac{UE} position $\mathbf{p}_0$ is assumed to be known.
At time $k$, given the previous position estimate $\hat{\mathbf{p}}_{k-1}$, the current position of the \ac{UE} $\hat{\mathbf{p}}_{k}$ is estimated.
Since the transmit power $P$ and pilot symbol phase may not be known at the \ac{BS}, a metric that is invariant to complex scaling is desirable.
For a given measurement $\mathbf{z}_{k,1:M}$, the position estimate at time $k$ is therefore obtained by maximizing the cosine similarity between the received signal and the predicted channel as~\cite{10345492}
\begin{equation}
\hat{\mathbf{p}}_k = \underset{{\mathbf{p}\in\mathcal{R}_k}}{\operatorname{argmax}} \sum_{m=1}^{M} \frac{\bigl\lvert\hat{\mathbf{h}}_{m}^{\mathrm{H}}(\mathbf{p}\!:\!\mathcal{S}_{\mathrm{a}})\mathbf{z}_{k,m}\bigr\rvert}{\big\|\hat{\mathbf{h}}_{m}(\mathbf{p}\!:\!\mathcal{S}_{\mathrm{a}})\big\|\big\|\mathbf{z}_{k,m}\big\|}\label{ob}
\end{equation}
where $\mathcal{R}_k$ denotes the search space at time $k$, constructed from the previous position estimate $\hat{\mathbf{p}}_{k-1}$ and \ac{UE} mobility constraints, as detailed in Sec.~\ref{PM}.
Since $\phi_{k,m}$ may differ across subcarriers, the cosine similarity is computed per subcarrier and its magnitude is summed, making each term individually invariant to the subcarrier-specific phase.

The predicted channel $\hat{\mathbf{h}}_{m}(\mathbf{p}\!:\!\mathcal{S}_{\mathrm{a}})$ inherently accounts for partial blockage and reflection effects determined by the known environment geometry, enabling accurate tracking under severe \ac{NLOS} conditions.
When $\eta<1$, surfaces in $\mathcal{S}\setminus\mathcal{S}_{\mathrm{a}}$ are invisible to the \ac{BS}: their blockage effects cause unmodeled path attenuation, and their reflection paths are absent from the predicted channel.
Both effects degrade the cosine similarity at the true \ac{UE} position, leading to increased tracking error.

\section{Environment-aware UE Tracking}\label{PM}
This section presents the proposed method for solving~\eqref{ob}.
Rather than relying on Kalman or particle filters, which require linearization that can introduce substantial model mismatch in the highly nonlinear \ac{NF} channel under partial blockage, this paper directly evaluates the full channel vector at each candidate position.
The approach involves two steps: constructing a tractable search space $\mathcal{R}_k$ from \ac{UE} mobility constraints, and computing the predicted channel $\hat{\mathbf{h}}_{m}(\mathbf{p}\!:\!\mathcal{S}_{\mathrm{a}})$ to maximize the cosine similarity.

\subsection{Search Space Construction}

The search space $\mathcal{R}_k$ at time $k$ is constructed based on \ac{UE} mobility constraints and the previous position estimate $\hat{\mathbf{p}}_{k-1}$.
Assuming the \ac{UE} moves with a maximum speed $v_{\max}$, the search space is defined as
\begin{equation}
\mathcal{R}_k =\left\{\mathbf{p}\,:\,\lVert\mathbf{p}-\hat{\mathbf{p}}_{k-1}\rVert\le v_{\max}\Delta_{\mathrm{t}}+E\right\}\label{ps}
\end{equation}
where $\Delta_{\mathrm{t}}$ denotes the tracking time interval and $E$ represents a margin to account for potential tracking errors from the previous time step.
This formulation ensures that $\mathcal{R}_k$ encompasses all physically reachable positions given the \ac{UE} mobility constraints, while the margin $E$ provides robustness against accumulated estimation errors.

The search space $\mathcal{R}_k$ is discretized into a uniform grid $\mathcal{G}_k$ with spacing $\Delta_{\mathrm{g}}$, where the grid spacing $\Delta_{\mathrm{g}}$ controls the trade-off between tracking accuracy and computational complexity.
Given the above, the position estimate $\hat{\mathbf{p}}_k$ is obtained as the grid point that maximizes the objective in~\eqref{ob} over $\mathcal{G}_k$.

\subsection{Environment-aware NF UE Tracking}\label{sec:ch_pred}

For each candidate grid point $\mathbf{p}\in\mathcal{G}_k$, the \ac{BS} constructs the predicted channel vector $\hat{\mathbf{h}}_{m}(\mathbf{p}\!:\!\mathcal{S}_{\mathrm{a}})$ defined in~\eqref{hhat} using only the known surface subset $\mathcal{S}_{\mathrm{a}}$.
The computation consists of three steps:
\begin{itemize}
\item \textbf{Step 1} (LOS path determination): For each antenna element $n$, evaluate the \ac{LOS} path indicator $b_{n}^{\mathrm{LOS}}(\mathbf{p}\!:\!\mathcal{S}_{\mathrm{a}})$ defined in~\eqref{blos} with $\mathcal{S}_{\mathrm{a}}$ in place of $\mathcal{S}$. 
The indicator equals $1$ if the line segment $\overline{\mathbf{p}_{n}'\mathbf{p}}$ does not intersect any surface in $\mathcal{S}_{\mathrm{a}}$, and $0$ otherwise.
\item \textbf{Step 2} (NLOS path determination): For each surface $s\in\mathcal{S}_{\mathrm{a}}$ and antenna element $n$, compute the reflection point $\mathbf{r}_{n,s}(\mathbf{p})$ via the image method and evaluate the \ac{NLOS} path indicator $b_{n,s}^{\mathrm{NLOS}}(\mathbf{p}\!:\!\mathcal{S}_{\mathrm{a}})$ defined in~\eqref{bnlos} with $\mathcal{S}_{\mathrm{a}}$ in place of $\mathcal{S}$.
\item \textbf{Step 3} (Channel vector construction): Assemble the predicted path set $\mathcal{L}_{n}(\mathbf{p}\!:\!\mathcal{S}_{\mathrm{a}})$ using~\eqref{pathset} with $\mathcal{S}_{\mathrm{a}}$, and compute $\hat{h}_{m,n}(\mathbf{p}\!:\!\mathcal{S}_{\mathrm{a}})$ via~\eqref{hhat} for all antenna elements and subcarriers.
\end{itemize}

After the three steps, the position estimate at time $k$ can be obtained by evaluating
\begin{equation}
\hat{\mathbf{p}}_k = \underset{{\mathbf{p}\in\mathcal{G}_k}}{\operatorname{argmax}} \sum_{m=1}^{M} \frac{\bigl\lvert\hat{\mathbf{h}}_{m}^{\mathrm{H}}(\mathbf{p}\!:\!\mathcal{S}_{\mathrm{a}})\mathbf{z}_{k,m}\bigr\rvert}{\big\|\hat{\mathbf{h}}_{m}(\mathbf{p}\!:\!\mathcal{S}_{\mathrm{a}})\big\|\big\|\mathbf{z}_{k,m}\big\|}\label{ob2}
\end{equation}
which is the grid-discretized counterpart of~\eqref{ob}.

Note that in contrast to the true channel vector $\mathbf{h}_{m}(\mathbf{p})$, which depends on all $S$ surfaces via $\mathcal{L}_{n}(\mathbf{p}\!:\!\mathcal{S})$, the predicted channel accounts for blockage and reflection only through the surfaces in $\mathcal{S}_{\mathrm{a}}$.
Surfaces in $\mathcal{S}\setminus\mathcal{S}_{\mathrm{a}}$ are invisible to the \ac{BS}: their blockage effects on existing paths are unmodeled, and their reflection paths are absent from $\mathcal{L}_{n}(\mathbf{p}\!:\!\mathcal{S}_{\mathrm{a}})$.

\section{Simulation Results and Case Studies}\label{sec:case}

\begin{figure*}[ht]
\vspace{-2.5ex}
\centering
\subfigure[\Ac{NF} tracking ($\eta=1$).]
{
\begin{tikzpicture}
\node[anchor=south west,inner sep=0] (img) at (0,0)
{\includegraphics[width=.3\linewidth]{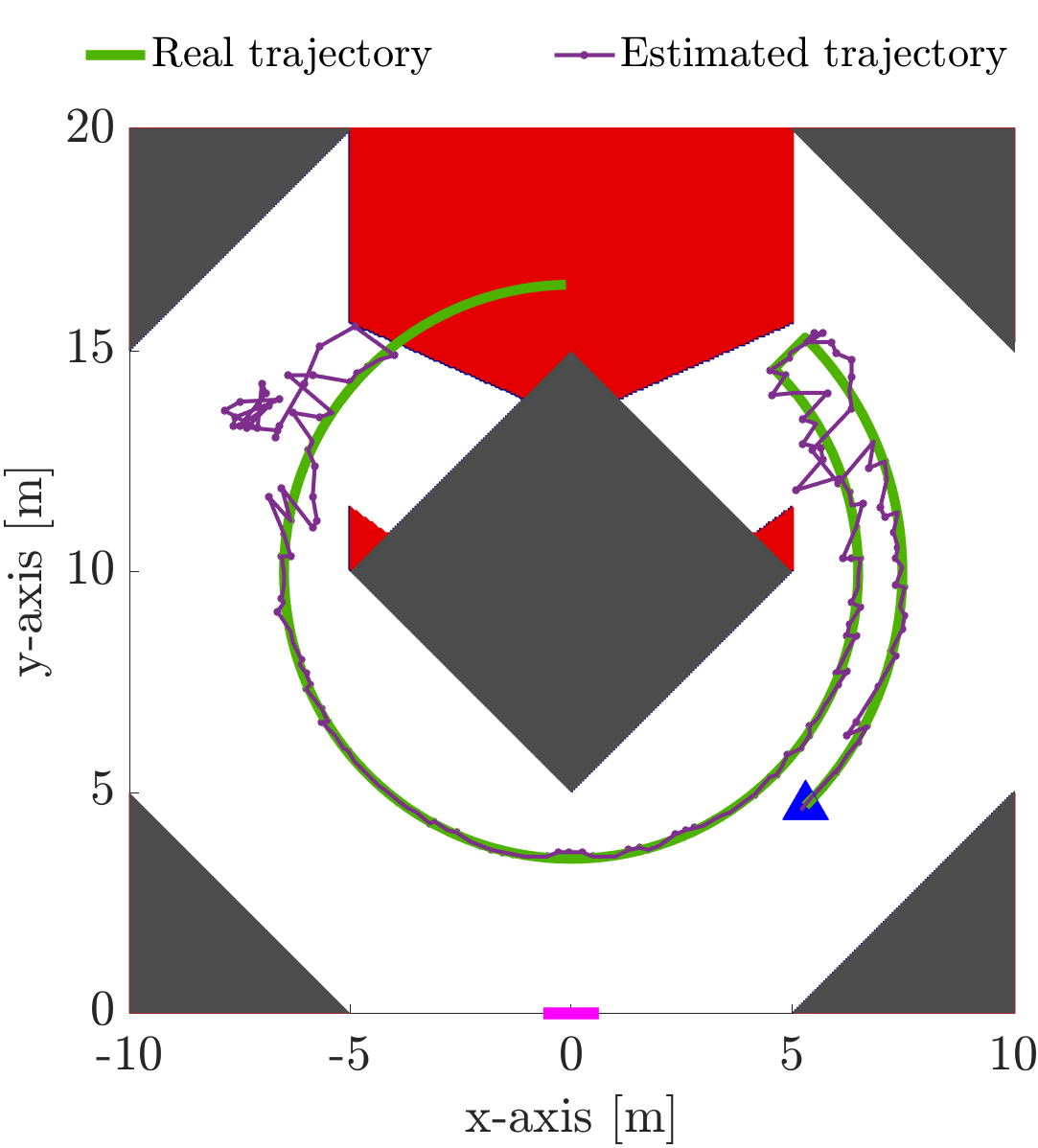}};
\begin{scope}[x={(img.south east)},y={(img.north west)},
every node/.style={font=\scriptsize\bfseries,
fill=white, inner sep=1pt, rounded corners=1pt,
opacity=0.85, text opacity=1}]
\node at (0.87,0.35) {\ding{192}};   
\node at (0.87,0.67) {\ding{193}};   
\node at (0.75,0.40) {\ding{194}};   
\node at (0.55,0.29) {\ding{195}};   
\node at (0.35,0.40) {\ding{196}};   
\node at (0.35,0.62) {\ding{197}};   
\node at (0.55,0.79) {\ding{198}};   
\end{scope}
\end{tikzpicture}
\label{trn212}
}
\subfigure[\Ac{NF} tracking ($\eta=0$).]
{
\includegraphics[width=.3\linewidth]{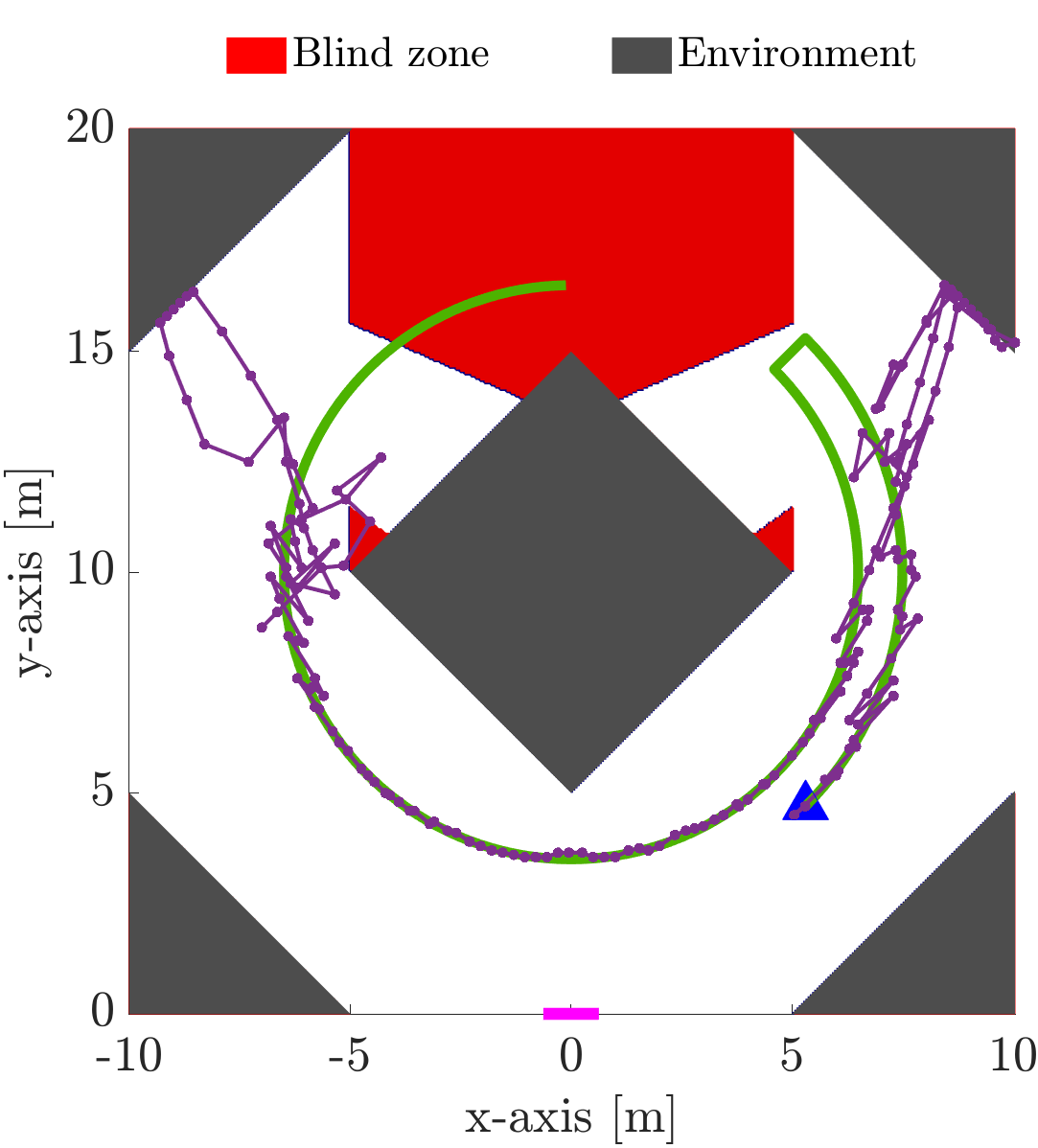}
\label{trn000}
}
\subfigure[\Ac{FF} tracking ($\eta=1$).]
{
\includegraphics[width=.3\linewidth]{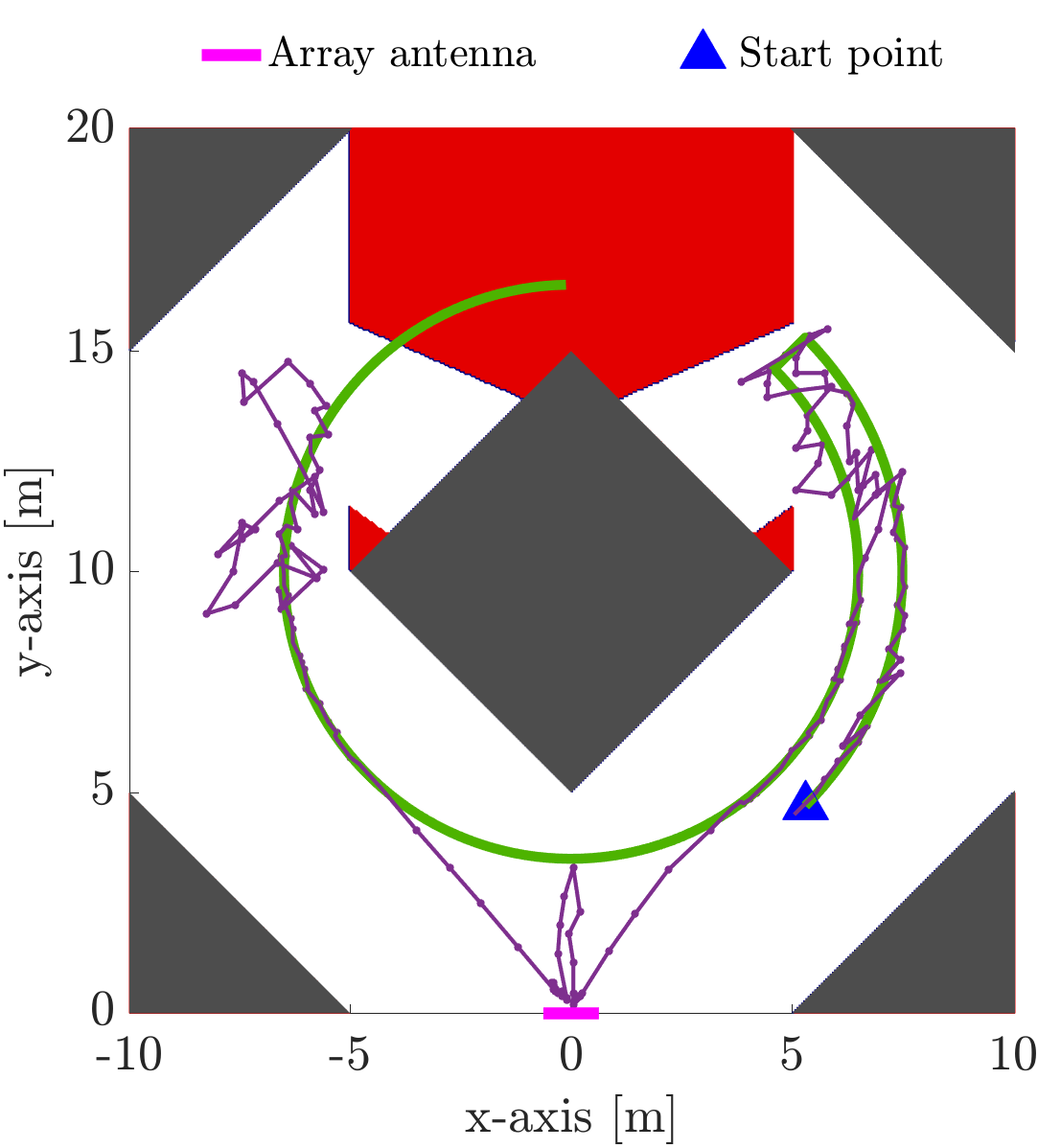}
\label{trf212}
}
\vspace{-1.5ex}
\caption{Tracking trajectories under three representative conditions.
The \ac{UE} traverses propagation zones
\ding{192}~\ac{LOS}+\ac{NLOS},
\ding{193}~\ac{NLOS},
\ding{194}~\ac{LOS}+\ac{NLOS},
\ding{195}~\ac{LOS},
\ding{196}~\ac{LOS}+\ac{NLOS},
\ding{197}~\ac{NLOS}, and
\ding{198}~blind zone, in order.}
\label{trs}
\vspace{-1.5ex}
\end{figure*}
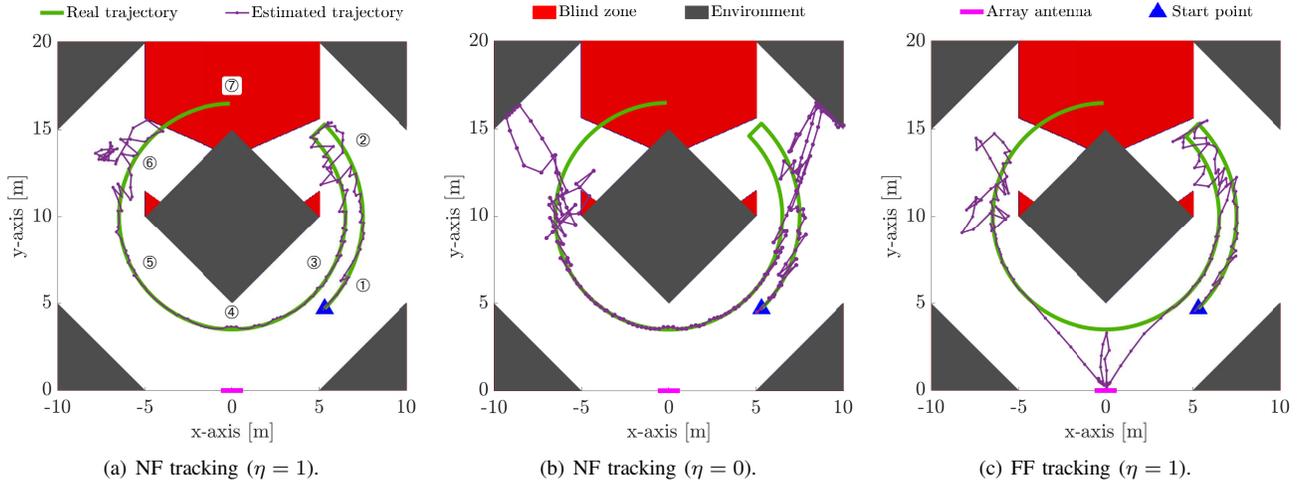

In this section, the proposed environment-aware \ac{NF} \ac{UE} tracking method is evaluated through numerical simulations.
The simulated environment consists of $S=8$ surfaces that act as reflectors and blockers under single-bounce propagation, as illustrated by the grey areas in Fig.~\ref{trs}.
For each environment-awareness level $\eta$, the known subset $\mathcal{S}_{\mathrm{a}}$ is drawn uniformly at random from $\mathcal{S}$, and the tracking \ac{RMSE} is averaged over multiple realizations and all trackable \ac{UE} positions.
Further simulation parameters are summarized in Tab.~\ref{table:sim_params}.
\begin{table}[hb]\vspace{-3.5ex}
\rowcolors{1}{white}{blue!10!white}
\centering
\caption{Simulation parameters.
\label{table:sim_params}}\vspace{-0.5ex}
    \resizebox{0.95\linewidth}{!}{
    \begin{tabular}{|l|l|c|}
    \hline
    \textbf{Description} & \textbf{Parameter} & \textbf{Value}\\ \hline
    Number of antennas          & $N$                  & $64$ \\
    Number of surfaces          & $S$                  & $8$ \\
    Carrier frequency           & $f_c$                & $7.5\,\mathrm{GHz}$ \\
    Bandwidth                   & -                    & $100\,\mathrm{MHz}$ \\
    Subcarrier spacing          & $\Delta_{\mathrm{f}}$& $120\,\mathrm{kHz}$ \\
    Number of subcarriers for tracking & $M$                  & $16$ \\
    Transmit power              & $P$                  & $23\,\mathrm{dBm}$ \\
    Noise figure                & $F$                  & $7\,\mathrm{dB}$ \\
    Kelvin temperature          & $T$                  & $290\,\mathrm{K}$ \\
    Tracking interval           & $\Delta_{\mathrm{t}}$           & $0.05\,\mathrm{s}$ \\
    Maximum speed               & $v_{\max}$           & $10\,\mathrm{m/s}$ \\
    Error margin                & $E$                  & $0.5\,\mathrm{m}$ \\
    Grid spacing                & $\Delta_{\mathrm{g}}$& $0.05\,\mathrm{m}$ \\
    \hline
    \end{tabular}
    }
\end{table}

\begin{figure*}[ht]
\vspace{1ex}
\centering
\subfigure[\Ac{NF} tracking ($\eta=1$).]
{
\includegraphics[width=.3\linewidth]{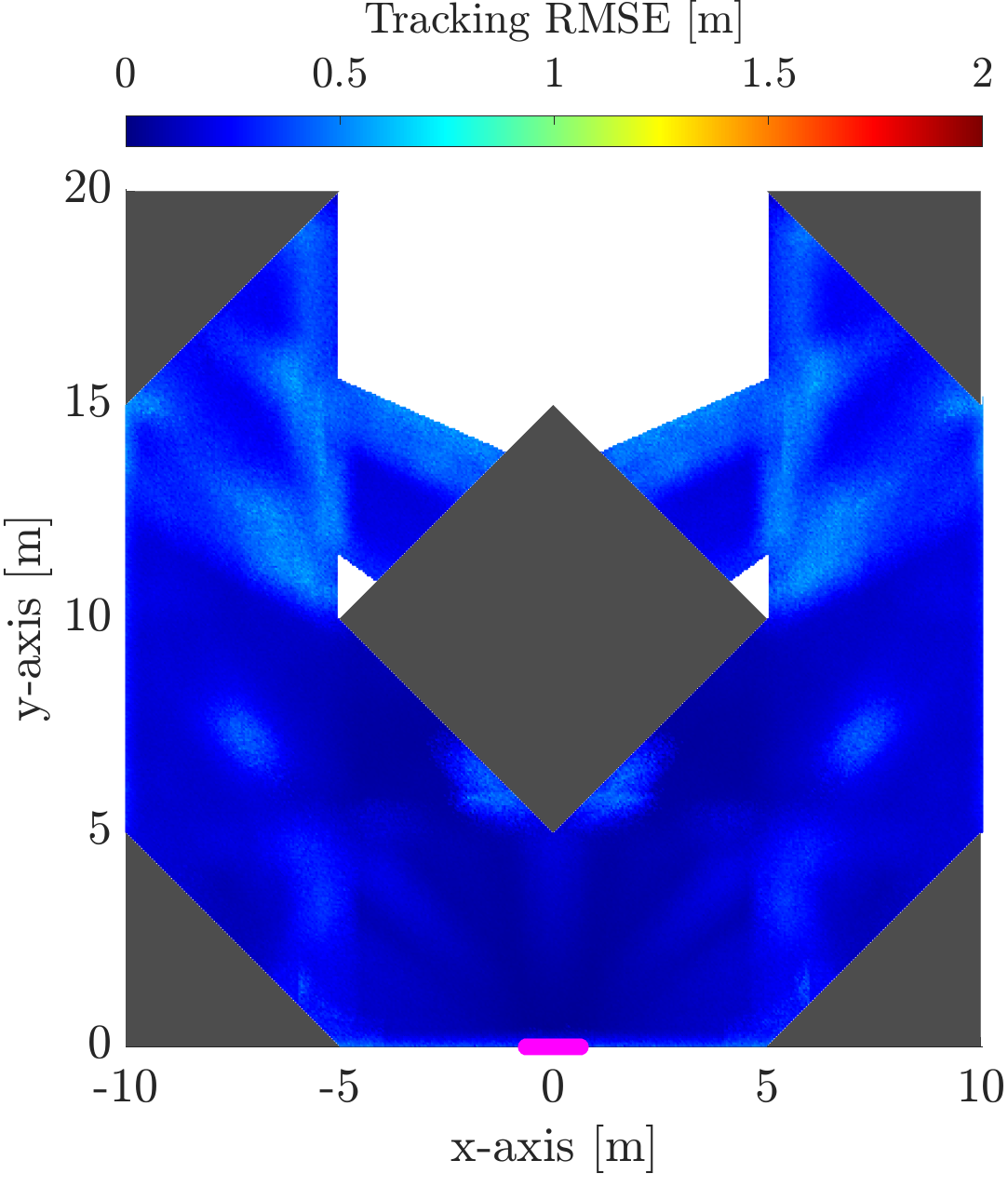}
\label{n212}
}
\subfigure[\Ac{NF} tracking ($\eta=0$).]
{
\includegraphics[width=.3\linewidth]{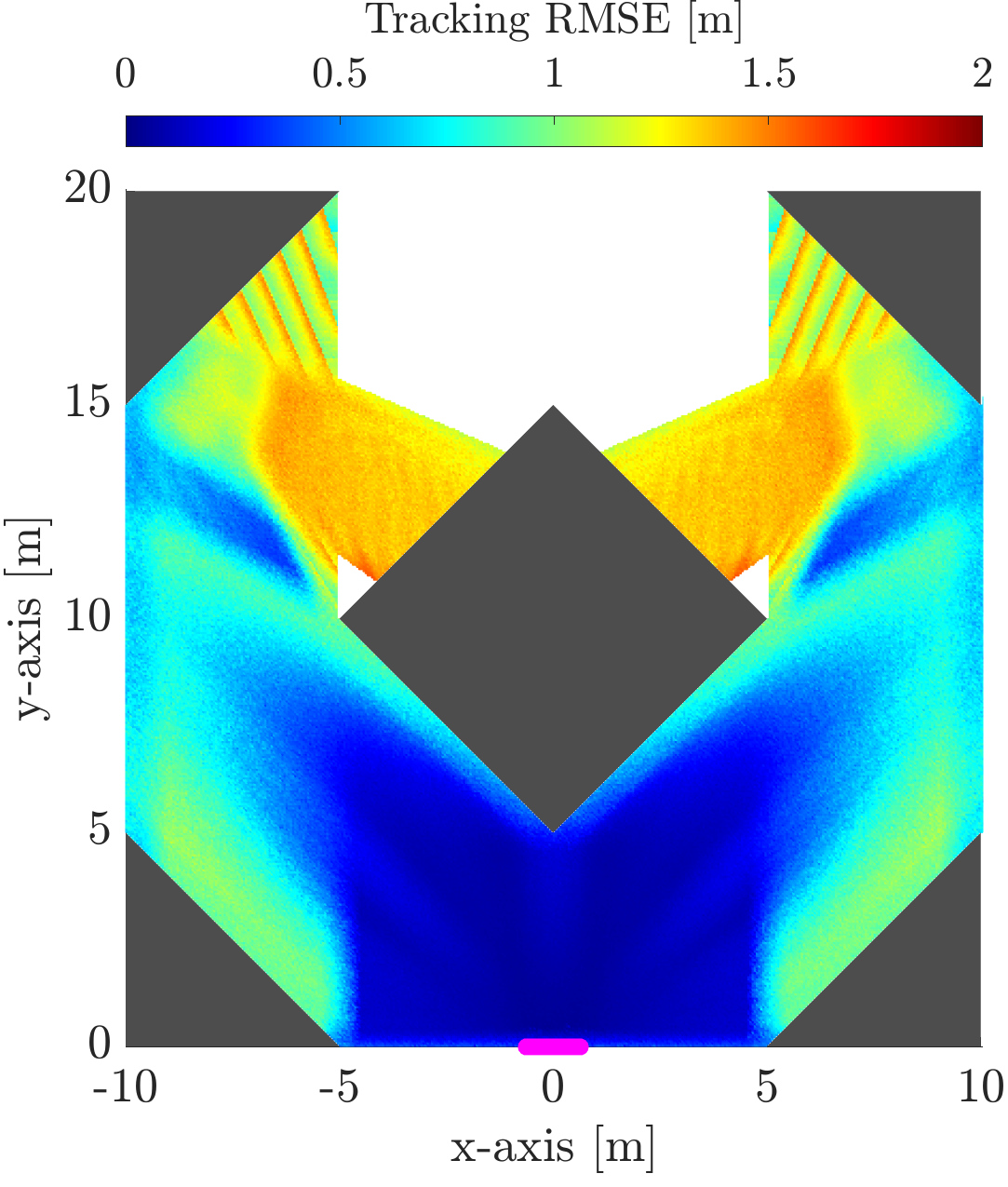}
\label{n000}
}
\subfigure[\Ac{FF} tracking ($\eta=1$).]
{
\includegraphics[width=.3\linewidth]{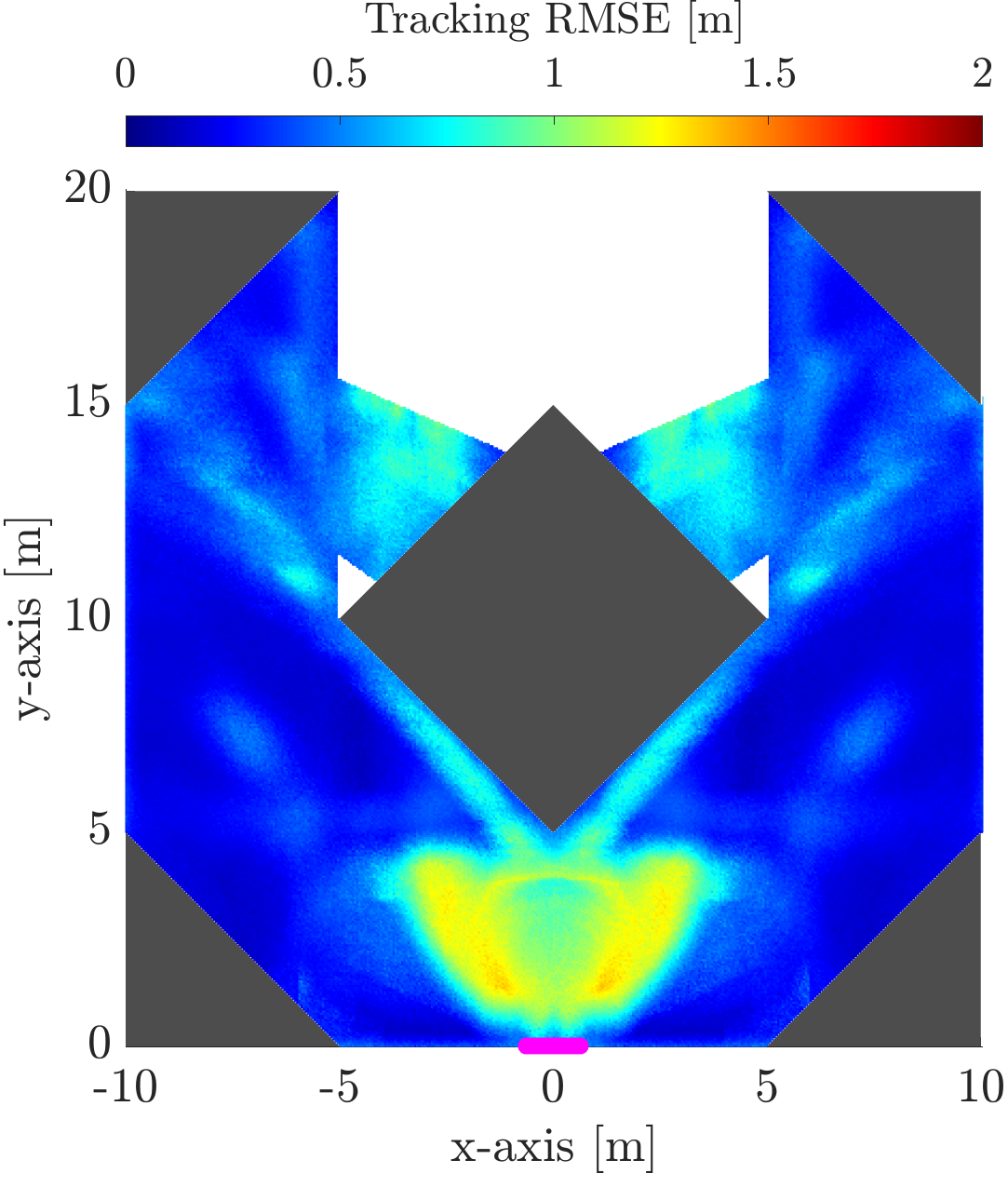}
\label{f212}
}
\vspace{-1.5ex}
\caption{Spatial distribution of tracking \ac{RMSE} over the trackable area.}
\label{rmses}
\vspace{-3ex}
\end{figure*}

Two baselines are adopted for comparison.
First, to isolate the gain from environment-awareness, an \ac{NF} variant that performs grid search using only the \ac{LOS} channel without any surface knowledge (i.e., $\eta=0$) is considered, following the framework in~\cite{10345492}.
Second, to isolate the gain attributable to the \ac{NF} model, a \ac{FF} variant that shares the identical environment-aware framework but replaces per-element path indicators with array-common ones is adopted, i.e., $b_{n}^{\mathrm{LOS}}(\mathbf{p}\!:\!\mathcal{S}_{\mathrm{a}})$ and $b_{n,s}^{\mathrm{NLOS}}(\mathbf{p}\!:\!\mathcal{S}_{\mathrm{a}})$ are constant across $n$, so that partial blockage and partial reflection effects are not captured.

Accordingly, Fig.~\ref{trs} illustrates the tracking trajectories for the proposed method and the two baselines.
With full environment knowledge under the \ac{NF} model (Fig.~\ref{trn212}), the estimated trajectory closely follows the true path throughout the entire course, including regions where the \ac{LOS} is completely blocked (excluding the complete blind zone), since the \ac{BS} can predict reflection paths from the known surfaces and exploit them for position estimation.
Without environment knowledge (Fig.~\ref{trn000}), the tracker relies solely on the \ac{LOS} path for channel prediction; the estimated trajectory remains accurate in \ac{LOS}-dominant regions, but degrades noticeably as the \ac{UE} enters \ac{NLOS}-dominant areas and fails when the \ac{LOS} is absent.
Comparing Figs.~\ref{trn212} and~\ref{trf212} reveals the limitation of the \ac{FF} assumption.
Although full environment knowledge is available in both cases, the \ac{FF} tracker (Fig.~\ref{trf212}) exhibits performance degradation at positions close to the \ac{BS} and at locations where partial blockage or partial reflection occurs.

Since the planar-wavefront model assigns identical path indicators across all antenna elements, it cannot capture the spatially-selective propagation phenomena that arise under the spherical wavefront, leading to a mismatch between the predicted and true channel vectors at these positions.

In complement, Fig.~\ref{rmses} presents the spatial distribution of the tracking \ac{RMSE} over \textit{all} trackable positions.
Under the proposed \ac{NF} method with $\eta=1$ (Fig.~\ref{n212}), the \ac{RMSE} remains below $0.45\,\mathrm{m}$ across the entire trackable area.
When $\eta=0$ (Fig.~\ref{n000}), comparable accuracy is maintained in \ac{LOS}-only regions, but the \ac{RMSE} increases progressively as the \ac{UE} enters \ac{NLOS}-dominant regions where unmodeled multipath distorts the cosine similarity metric.
For the \ac{FF} baseline with $\eta=1$ (Fig.~\ref{f212}), elevated \ac{RMSE} values persist near the \ac{BS} and at partial blockage/reflection positions, where the all-or-nothing path model deviates from the true propagation.

\begin{figure}[t]
\centering
\includegraphics[width=0.95\linewidth]{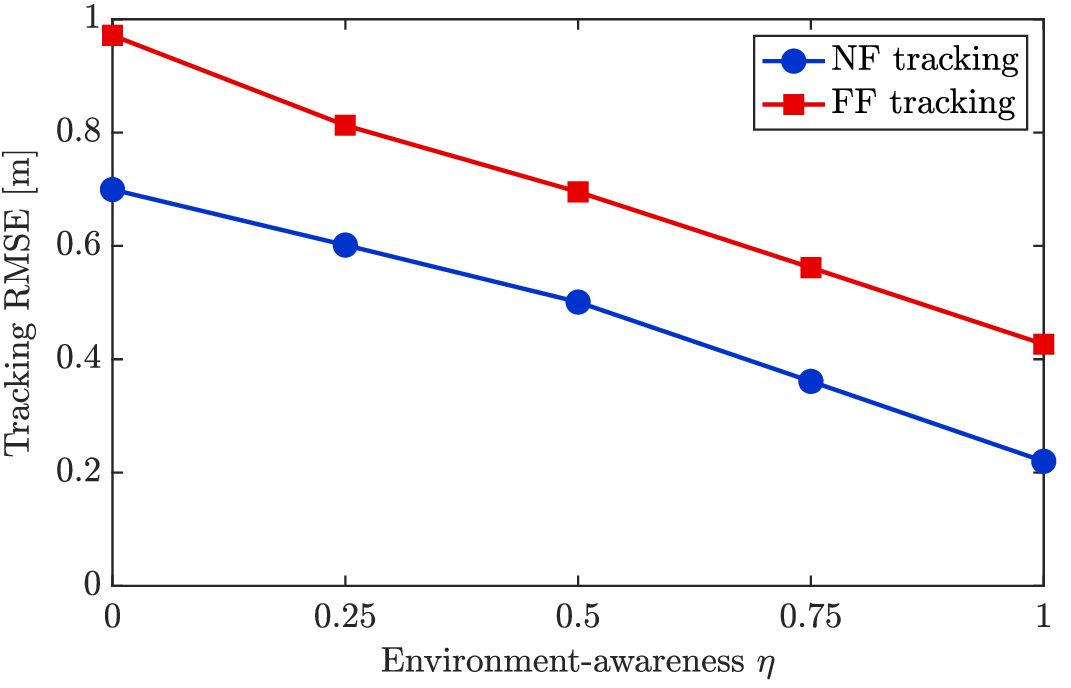}
\vspace{-2ex}
\caption{Tracking \ac{RMSE} versus environment-awareness level $\eta$.}
\label{rmsenf}
\vspace{-2.5ex}
\end{figure}

Finally, Fig.~\ref{rmsenf} summarizes the average tracking \ac{RMSE} as a function of the environment-awareness level $\eta$, from 0 to 1, for both the proposed \ac{NF} and the \ac{FF} methods.
Both methods benefit from increasing environment-awareness, as additional surface knowledge enables more accurate channel prediction.
In particular, the proposed \ac{NF} method reduces the average \ac{RMSE} from $0.70\,\mathrm{m}$ at $\eta=0$ to $0.22\,\mathrm{m}$ at $\eta=1$.
Moreover, the proposed \ac{NF} method consistently outperforms the \ac{FF} baseline across all values of $\eta$.
The performance gap persists even at $\eta=1$, confirming that the advantage of the \ac{NF} model stems not only from environment-awareness but also from the accurate modeling of partial blockage and partial reflection.
These results demonstrate that incorporating both environment geometry and \ac{NF} propagation characteristics into the tracking framework is essential for achieving high-accuracy \ac{UE} tracking in \ac{ELAA} systems.

For each candidate position $\mathbf{p}\in\mathcal{G}_k$, evaluating the \ac{LOS} and \ac{NLOS} indicators requires $\mathcal{O}(NS_{\mathrm{a}}^2)$ intersection checks, and constructing the channel vector across $M$ subcarriers costs $\mathcal{O}(NM(S_{\mathrm{a}}+1))$.
With $|\mathcal{G}_k|\approx(v_{\max}\Delta_{\mathrm{t}}+E)^2/\Delta_{\mathrm{g}}^2$ grid points, the total per-step complexity is $\mathcal{O}\bigl(|\mathcal{G}_k|\,N(S_{\mathrm{a}}^2+MS_{\mathrm{a}})\bigr)$.
Under the simulation parameters, the measured per-step execution time is less than $0.02\,\mathrm{s}$, which is well below the tracking interval $\Delta_{\mathrm{t}}=0.05\,\mathrm{s}$, confirming real-time feasibility.
The grid spacing $\Delta_{\mathrm{g}}=0.05\,\mathrm{m}$ is chosen to be sufficiently fine so that it does not bottleneck the tracking accuracy.
\vspace{-1ex}
\section{Conclusion}\label{sec:conc}
This paper proposed an environment-aware \ac{NF} \ac{UE} tracking method that leverages known surface geometries to construct per-antenna-element channel predictions, thereby capturing partial blockage and partial reflection effects unique to the \ac{NF} regime.
By directly maximizing the cosine similarity between predicted and received channel vectors, the proposed method bypasses conventional channel parameter extraction and enables continuous tracking even under complete \ac{LOS} obstruction.
Numerical results confirmed that increasing environment-awareness progressively enhances tracking performance, and that the \ac{NF} model consistently outperforms the \ac{FF} baseline across all awareness levels, demonstrating the necessity of jointly exploiting environment geometry and \ac{NF} propagation characteristics for high-accuracy \ac{UE} tracking in \ac{ELAA} systems.
Future work includes extensions to \ac{3D} multi-bounce environments, multi-static placements of arrays, and robustness analysis under surface geometry uncertainties.

\end{document}